\documentclass[preprints,article,accept,moreauthors,pdftex]{Definitions/mdpi}

\usepackage{comment}
\usepackage{amsmath}
\usepackage{xcolor}
\firstpage{1} 
\pubvolume{1}
\issuenum{1}
\articlenumber{0}
\pubyear{2021}
\copyrightyear{2020}
\datereceived{} 
\dateaccepted{} 
\datepublished{} 

\Title{Gate control of superconductivity in mesoscopic all-metallic devices}

\TitleCitation{Electrostatic control of superconductivity in mesoscopic all-metallic devices}

\Author{Claudio Puglia $^{1,2,^*}$\orcidA{}, Giorgio De Simoni $^{2}$\orcidB{} and Francesco Giazotto $^{2}$\orcidC{}}

\AuthorNames{Claudio Puglia, Giorgio De Simoni and Francesco Giazotto}

\AuthorCitation{Puglia, C.; De Simoni, G.; Giazotto, F.}

\address{%
$^{1}$ \quad Dipartimento di Fisica dell’Università di Pisa, Largo Pontecorvo 3, I-56127 Pisa, Italy\\
$^{2}$ \quad NEST, Instituto Nanoscienze-CNR and Scuola Normale Superiore, I-56127 Pisa, Italy}

\corres{Correspondence: claudio.puglia@df.unipi.it}

\abstract{It was recently demonstrated the possibility to tune, through the application of a control gate voltage, the superconducting properties of mesoscopic devices based on Bardeen-Cooper-Schrieffer metals. In spite of the several experimental evidence obtained on different materials and geometries, a description of the microscopic mechanism at the basis of such unconventional effect has not been provided yet. This work discusses the technological potential of gate control of superconductivity in metallic superconductors and revises the experimental results which provide information regarding a possible thermal origin of the effect: in the first place, we review experiments performed on high critical temperature elemental superconductors (niobium and vanadium) and show how devices based on these materials can be exploited to realize basic electronic tools such as, e. g., a half-wave rectifier. In a second part, we discuss the origin of the gating effect by showing the gate-driven suppression of the supercurrent in a suspended titanium wire and by providing a comparison between thermal and electric switching current probability distributions. Furthermore, we discuss the cold field-emission of electrons from the gate by means of finite element simulations and compare the results with experimental data. Finally, the presented data provide a strong indication regarding the unlikelihood of thermal origin of the gating effect.}


\begin{document}

\section{Introduction}
In the last two years, the impact of gate voltage on the superconducting properties of Bardeen-Cooper-Schrieffer (BCS) \cite{Bardeen1962} metallic superconductors has been investigated \cite{DeSimoni2018,Paolucci2018,Paolucci2019,Paolucci2019b,Paolucci2019a,Bours2020}. In these studies the authors analyze the effect of the electrostatic gating, generating an electrostatic field of the order of $10^8$ V/m and, at the same time, modifying negligibly the concentration of the surface charge carriers. Particularly, ampbipolar suppression of supercurrent has been demonstrated in all-metallic superconductor wires \cite{DeSimoni2018}, nano-constriction Josephson junctions (JJs) \cite{Paolucci2018,Paolucci2019}, fully-metallic Superconducting Quantum Interference Devices (SQUID) \cite{Paolucci2019a}, and proximity nanojunctions \cite{DeSimoni2019}. Such unconventional gating effect in BCS superconductor systems is the first step for the realization of easy fabrication and high-scalable technologies in both the environments of classical and quantum electronics. The aim of this work is to review the most recent advances on the modification of superconducting properties in mesoscopic structures via the application of a control gate voltage. For such an effect, a fulfilling microscopic theory has not been provided yet. Indeed, it is not possible to take into account for experimental observations through the conventional BCS framework, in which the superconducting properties are negligibly affected by electric fields \cite{Virtanen2019}. Although, some theories have been proposed, including the electric field driven Rashba orbital polarization \cite{Mercaldo2020}, and the gate-driven Schwinger excitation of quasiparticles from the BCS vacuum \cite{Solinas2020}, they have not been experimentally verified yet. The injection of high-energy field-emitted cold-electrons into the weak-link was also hypothesized to be at the origin of the gating effect \cite{Alegria2021,Ritter2020}. Nevertheless, even in presence of the latter mechanism, several experimental results are not compatible with a mere power injection resulting in an overeating of the superconductor \cite{DeSimoni2018,Paolucci2018,Puglia2020,Rocci2020}.

The article is organized as follows: Section \ref{sec:mat} displays evidence of a gate-driven supercurrent suppression in vanadium and niobium Dayem bridges (DBs). Moreover, different technological implementations based on these materials are presented. In the same section, two further topics are faced. The former is the modification of the switching current probability distribution as a function of the electric field. The latter is the influence of the substrate on the gating effect in titanium weak-links. Section \ref{sec:thermal} analyses the  evidence against a thermal origin of the supercurrent suppression. Finally, Section \ref{sec:sum} provides a summary of the results presented in this review resuming the main achievements and proposing new experiments to increase the understanding of the gating effect.

\section{Gate-driven supercurrent suppression in Nb and V nanojunctions}\label{sec:mat}

In this section we present a series of experiments, performed on niobium and vanadium superconducting weak-link devices, aimed at extending the range of materials suitable for gated-superconductor applications at elemental superconductors with critical temperature higher than the liquid helium temperature $\sim4.2$ K. The presented results demonstrate the possibility to implement gate-controlled all-metallic superconducting electronics \cite{Likharev2012} compatible with industrial standards.

\subsection{Niobium gate-controlled transistor}
\begin{figure}[ht!]
\includegraphics[width=13.4 cm]{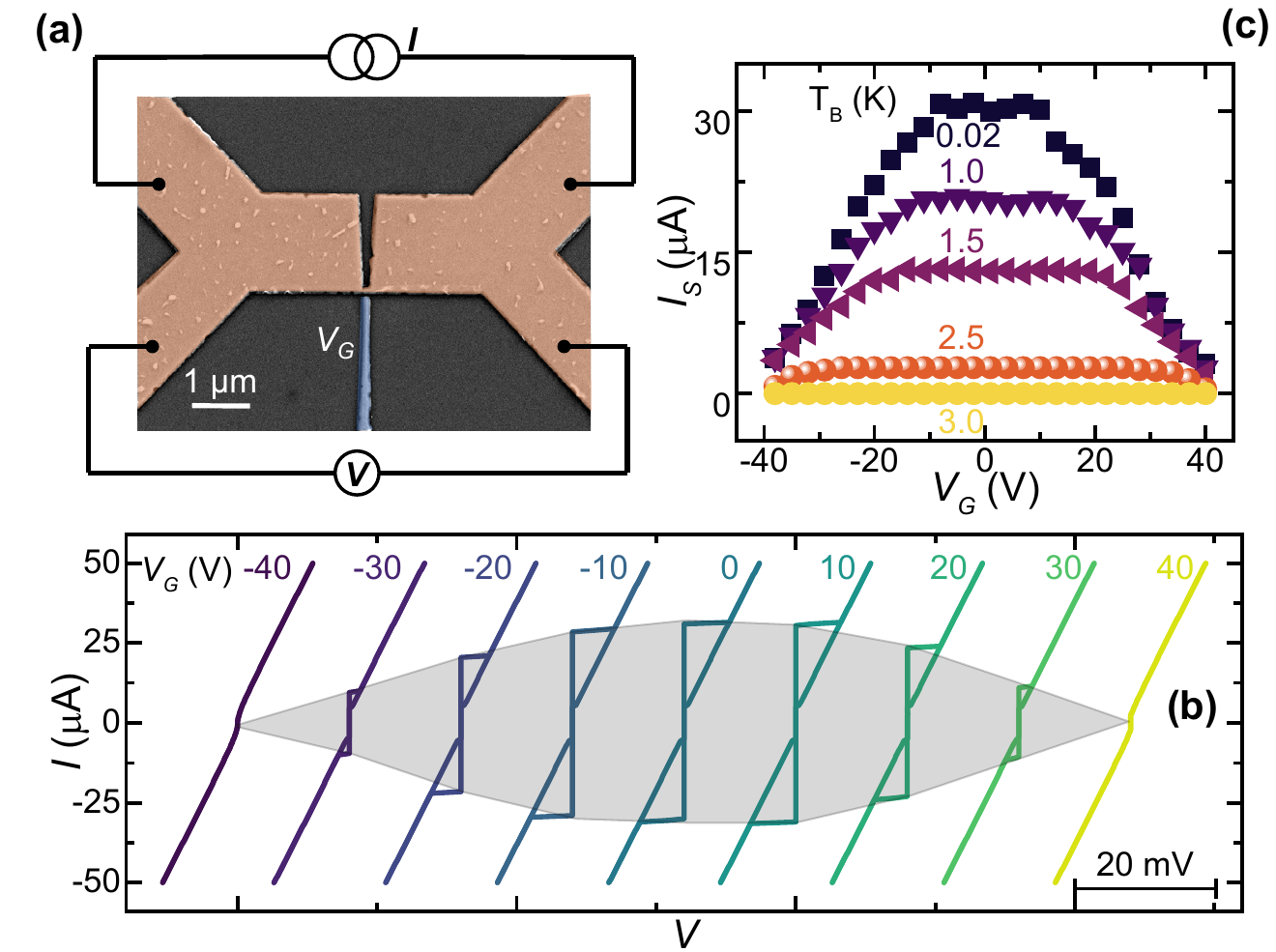}
\caption{ \textbf{(a)} Pseudo-color SEM of a representative niobium gate-tunable transistor with the four-probes bias scheme. The junction and the wire are in false-colored orange, the gate in blue. \textbf{(b)} $I$ vs $V$ curves for different gate voltages $V_G$ at a bath temperature of 20 mK. Characteristics are horizontally shifted for clarity. A clear symmetric suppression of $I_S$ is visible as the modulus of the gate voltage $|V_G|$ is increased. \textbf{(c)} $I_S$ vs $V_G$ for several bath temperatures $T$ in a range from $20$ mK to $3$ K. The switching current $I_S$ values were acquired by acquiring $50$ repetitions of the $I$ vs $V$ characteristics. Error bars are smaller than the data symbol.}
\label{fig:NbDev}
\end{figure}

All-metallic supercurrent transistors consist of a superconducting mesoscopic channel realized with BCS metals, equipped with gate electrodes lithographically fabricated at a distance of a few nanometer from the channel. The gate electrode is polarized through the application of an either positive or negative control gate voltage. Niobium gate-controlled transistors typically consist of a 8-$\mu$m-long, 2.5-$\mu$m-wide wire interrupted by a 50-nm-wide, 120-nm-long constriction. Aligned with the DB weak-link it a was realized a co-planar, 60-nm-far, 80-nm-wide metallic gate. The thin film was deposited on a sapphire Al$_2$O$_3$ substrate via DC magneto-sputter deposition of a 10/40-nm-thick Ti/Nb bylayer. The former metal was necessary to increase the adhesion and the mechanical strength of the metallic film. A pseudo-color scanning electron micrograph (SEM) is shown in Figure~\ref{fig:NbDev}(a).

The device shows a Dayem bridge normal state resistance $R_{DB} \simeq30\ \Omega$ and a critical temperature $T_{DB} \simeq3$ K \cite{DeSimoni2020}. On the other hand, the niobium banks inherit the critical temperature of the pristine thin film $T_{Nb}\simeq7.9$ K \cite{DeSimoni2020}. The smaller critical temperature of the Dayem bridge is due to its lateral size which is comparable with the niobium coherence length \cite{Stromberg1962,Finnemore1966}. $T_{Nb}$, instead, is about $80$\% of the conventional critical temperature for Nb because of the proximity effect of the adhesion titanium layer. The conventional dissipationless transport is displayed by plotting the current $I$ vs voltage $V$ curves acquired at a bath temperature $T = 20$ mK and a gate voltage $V_G=0$ V, as shown in Figure~\ref{fig:NbDev}(b). The Dayem bridge switching current is $I_S \simeq 30\ \mu$A. The forward and backward $I$ vs $V$ characteristics shows the characteristic hysteretic behavior due to the retrapping current $I_R$ \cite{Giazotto2006,Tinkham2004}.

The suppression of the switching current via the application of a gate voltage was demonstrated by measuring the current $I$ vs voltage $V$ characteristics with a gate voltage in the range from $-40$ to $40$ V at a bath temperature of $T=20$ mK. Figure~\ref{fig:NbDev}(b) displays $V(I)$ curves at several gate voltages. A shadow grey area is drawn to underline the suppression region. The quenching of the supercurrent is symmetric for $V_G\longrightarrow-V_G$ for bath temperatures between 20 mK and 3 K as shown in Figure~\ref{fig:NbDev}(c). As reported in conventional experiments \cite{DeSimoni2018,DeSimoni2019,Paolucci2018,Paolucci2019}, the widening of the plateau in which the gate voltage is not effective is clearly visible as the temperature rises. The suppression of the supercurrent can be observed up to a temperature of $3$ K with complete suppression at $|V_G|=40$ V for $T>2$ K.

\subsubsection{Rectification properties}

\begin{figure}[ht!]
\includegraphics[width=13.4 cm]{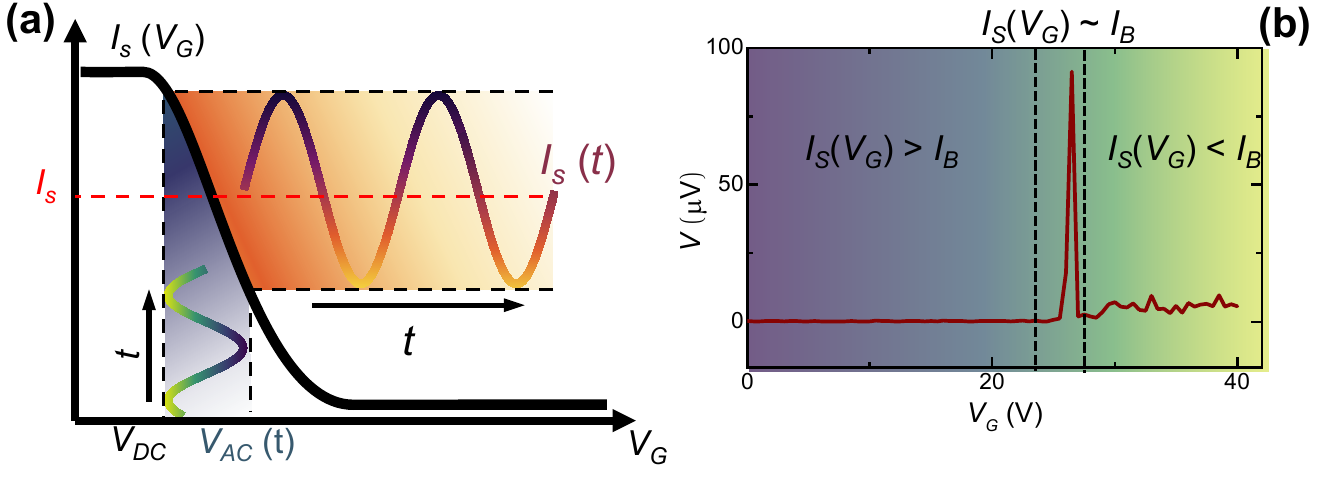}
\caption{\textbf{(a)} Sketch of the operation principle  exploited by the niobium-based rectifier. The system is DC current biased with $I_B$ (horizontal red dashed line in \textbf{(a)}) and the side-gate is carries a signal consisting of an alternate part $ V_{AC} $ (blue/green line) and a DC part $ V_{DC} $. The resulting characteristic $ I_S(t) $ (purple/yellow curve) depends on the amplitude of $ V_{AC} $ and on the value of $ V_{DC }$ and generates recurring super-to-normal and  normal-to-super state  transitions. In this way, the voltage signal across the junction $ V $ oscillates between a low and a high state with the same period of $V_{AC}$. \textbf{(b)} Voltage $ V $ acquired across constriction in a lock in four-wires scheme. The alternate voltage $ V_{AC} $ is the reference signal of the lock-in amplifier. The system is current biased with $I_B  =  2.5\ \mu$A. $V$ is almost zero until $I_S(V_G) < I_B$. The peak represent the rectification effect of the AC gate voltage signal. }
\label{fig:NbCorna}
\end{figure}

Basing on the peculiar shape of the $R$ vs $V_G$ characteristic \cite{DeSimoni2020}, it is possible to implement a superconducting diode scheme. In particular, an alternate gate voltage $V_{AC}$ can be rectified exploiting the sharpness of the super-to-normal transition when the junction is current biased. In this configuration, a sinusoidal gate voltage is transformed into a square wave voltage-drop across the junction. Such a peculiar system response is shown in Figure~\ref{fig:NbCorna}\textbf{(a)}. The gate voltage signal is the sum of $V_{AC}$ and of a DC pre-bias voltage used to define a switching current working range $I_S(V_G^{max})<I_B<I_S(V_G^{min})$. The oscillation of $I_S$ above and below $I_B$ results in time dependent normal-to-super and vice versa transitions that generate a time dependent voltage drop $V(t)$ across the junction as shown in Figure~\ref{fig:NbCorna}\textbf{(b)}. The output signal maintains the same periodicity of $V_{AC}$ and the duty cycle is defined by the time fraction for which the condition $I_S<I_B$ is satisfied. 

Notably, the output voltage depends directly on the amplitude of the AC input signal thanks to the behavior of $R$ vs $V_G$ characteristic. In the configuration shown in Figure~\ref{fig:NbCorna} our system acts as a half-wave rectifier which could be used in a vast range of devices such as diodes and detectors. In the next paragraph further  evidences of the rectification properties of such systems are provided, with emphases on the versatility of the technology.

\subsection{Vanadium gate-controlled transistor}
\begin{figure}[ht!]
\includegraphics[width=13.4 cm]{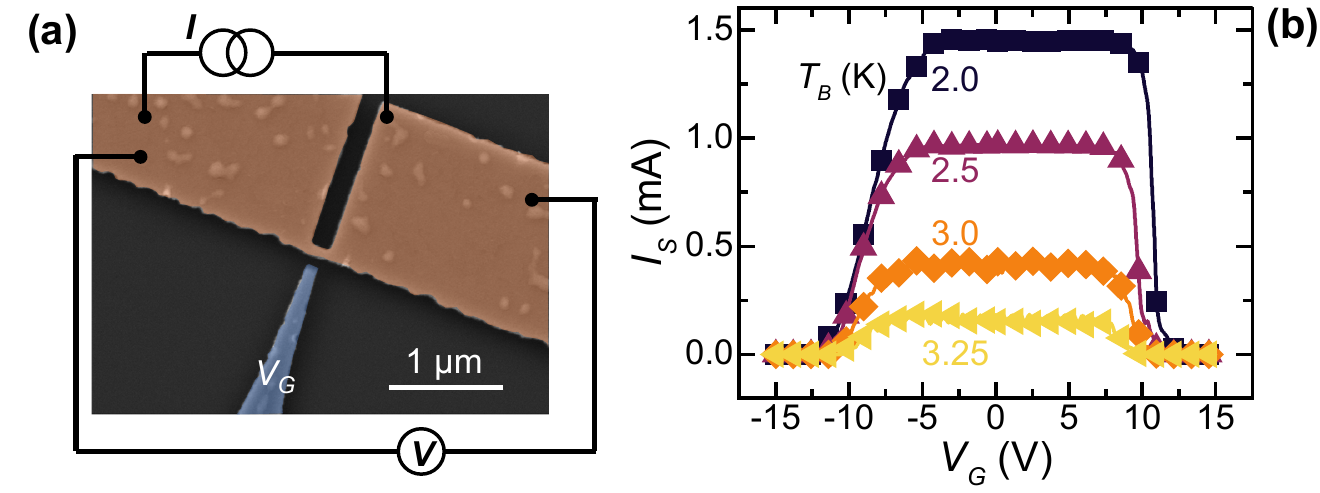}
\caption{\textbf{(a)} Pseudo-color SEM of a representative vanadium gate-tuned transistor. The junction and the wire are colored in orange, the gate in blue. \textbf{(b)} Current $I$ vs voltage $V$ curves of a typical device acquired at several bath temperatures from $ 2.0 $ to $ 3.8 $ K. The characteristics are horizontally shifted for clarity. \textbf{(c)} Switching current $I_S$ vs gate voltage $V_G$ curves for several bath temperatures in the range between $2.0$ and $3.3$ K. $I_S$ values were obtained averaging $25$ acquisitions.}
\label{fig:VDev}
\end{figure}

The vanadium gate-controlled transistors typically consist of a planar 60-nm-thick, 160-nm-long, 90-nm-wide Dayem bridge Josephson junction flanked by 70-nm-far, 120-nm-wide side-gate. Our exploited bridge geometry is similar to those already discussed for Nb-based devices. The system was realized by means of a one-step electron beam lithography performed on a silicon dioxide substrate (SiO$_2$) followed by the electron beam evaporation of high purity vanadium at a rate of 0.36 nm/s in an ultra-high vacuum chamber. Figure~\ref{fig:VDev} shows the pseudo-color SEM of a typical vanadium gate-controlled transistor.

The device shows a normal state resistance $R_{N} \simeq106\ \Omega$, a switching current at 2 K of $I_S=1.42$ mA and a critical temperature $T_{C} \simeq3.6$ K \cite{Puglia2020a} as displayed in Figure~\ref{fig:VDev}\textbf{(b)}.

The suppression of the supercurrent as a function of the gate voltage was demonstrated by measuring the $I_S$ vs $V_G$ characteristics. Figure~\ref{fig:VDev}\textbf{(c)} shows the bilateral suppression of the supercurrent down to total quenching for $|V_G|\simeq 10$ V in a range of bath temperature from $2$ to $3.2$ K. Notably, the sharper suppression of $I_S$ observed for positive values of the gate voltage is in contrast with a possible cold field-emission origin of the quenching effect. Indeed, the device geometry could facilitate the electron extraction from the gate that occurs at negative gate bias values \cite{Alegria2021,Ritter2020}. This consideration is deeply discussed in Section \ref{sec:thermal}.

\subsubsection{Half-wave rectifier}

\begin{figure}[!ht]
\centering
\includegraphics[width=13.4 cm]{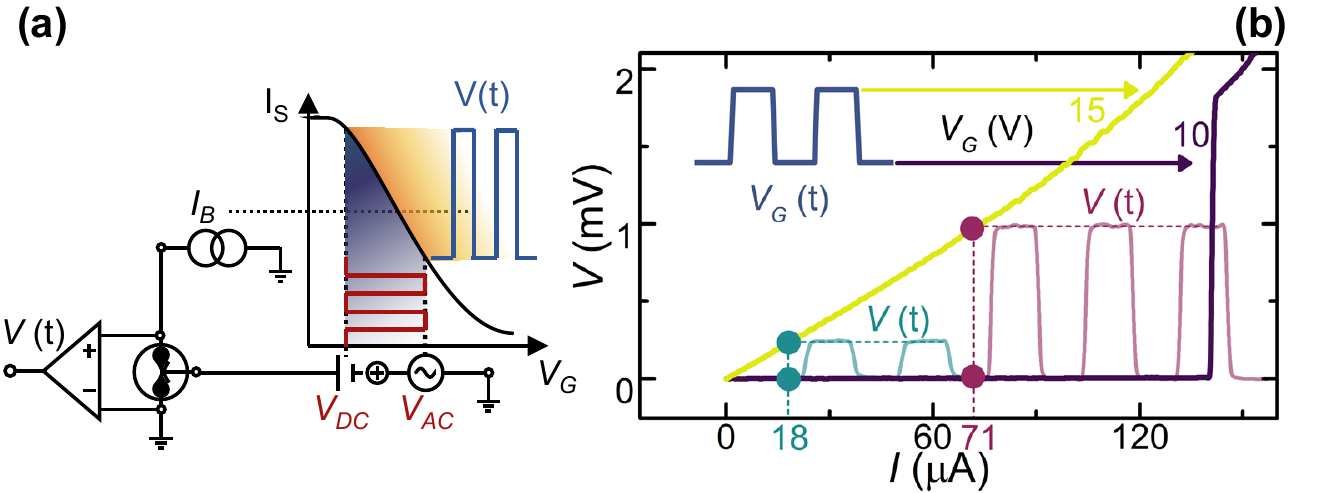}
\caption{\textbf{(a)} Bias scheme for AC measurements. The gate voltage is generated adding up a DC $V_{DC}$ and a AC $V_{AC}$ arbitrary waveform voltages. The ADC/CAD board that provides the AC signal performs a real-time measurements of the voltage drop $V$ across the junction. \textbf{(b)} Voltage $V$ vs current $I$ curves for chosen values of $V_G$. The two dot couples displays the operation  points of the system for two bias currents $I_B = 18$,  $71 \  \mu$A. The time dependent response of the system $V(t)$ is represented with the same color of the dot pairs. The gate voltage signal $V_G(t)$ consists of a DC voltage $ V_{ DC} = 10 $ V and an AC square-wave voltage with amplitude $ V_{AC} = 5 $ V. The bath temperature for these measurements was $ T = 3 $ K.}
\label{fig:tresponce}
\end{figure}

A time-resolved characterization of the device was carried out using both sinusoidal and square -wave gate voltages. Figure~\ref{fig:tresponce}\textbf{(a)} shows the bias scheme of the measurement setup consisting in a DC bias current, a DC voltage generator and an ADC/DAC digital board providing the AC gate voltage signal. The latter voltage generators provides a $V_G(t)=V_{DC}+V_{AC}(t)$ gate signal, setting the right operation point in the parameters space (see Figure~\ref{fig:tresponce}\textbf{(a)}).

The time dependent voltage drop $V(t)$ across the junction was acquired as a function of the time-dependent gate voltage $V_G(t)$ obtaining a zero-signal (low-state) when $I_S\left[V_G(t)\right]>I_B$ (superconducting state). By contrast, when $I_S\left[V_G(t)\right]<I_B$ the junction switches to the dissipative state and a voltage drop different from zero is obtained across the constriction (high-state). We measured the system response to a transistor–transistor logic-like (TTL) square-wave gate signal composed by a $V_{DC}=10$ V voltage DC bias plus a $V_{AC}=5$ V square-wave with a frequency up to $\sim50$ Hz, as shown in Figure~\ref{fig:tresponce}\textbf{(b)}.

The low and high states are highlighted on top of the $ I $ vs $ V $ characteristics (obtained for $ V_G $ ranging from $ 10 $ to $ 15 $ V) with couples of dots of the same colors in Figure~\ref{fig:tresponce}\textbf{(b)}. The $ V (t) $ signal resulting from the $V_G(t)$ excitation is shown in Figure~\ref{fig:tresponce}\textbf{(d)} for two current bias ($I_S=18 , \ 71\ \mu$A). We note that the output voltage is proportional to the bias current. It is worth to emphasize again that $ V (t) $ maintains the shape of the input voltage signal with frequencies, in principle, limited only by $ f_\Delta $ \cite{Barone1982,Tinkham2004}.

\begin{figure}[!ht]
\centering
\includegraphics[width=13.4 cm]{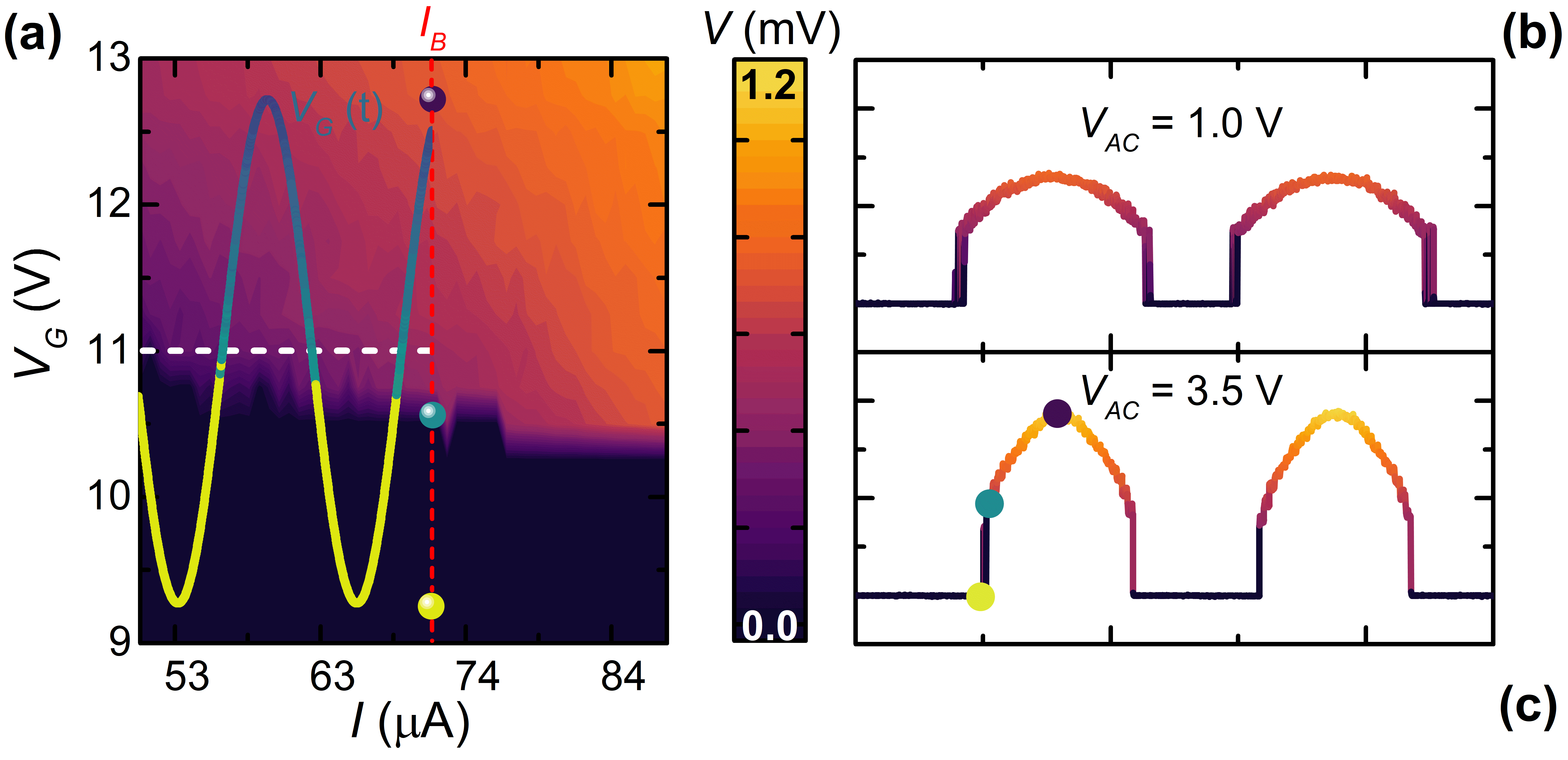}
\caption{\textbf{(a)}. Color-plot of the voltage drop $V(t)$ across the Dayem bridge constriction \textit{versus} the gate voltage $V_G$ (X-axis) and the bias current $I_B$ (Y-axis). The three dots represent the zero signal state (yellow), the transition to the normal-state (green) and the maximum $V(t)$ signal (blue), respectively. The vertical dashed red line shows the operating curve defined by the bias current. The yellow-to-blue sinusoidal represents the AC + DC gate voltage excitation $V_G(t)$.  \textbf{(b)} and \textbf{(c)} Time dependent voltage drop across the constriction for two amplitude of the alternate component of the gate voltage, $ V_{AC} = \  3.5,\  1.0 $ V. The color-map is all the panels. The measurements were performed at a bath temperature of $T = 3$ K.}
\label{fig:VSin}
\end{figure}

Finally, we show the response of the system to a sinusoidal gate voltage signal. The measurement setup is the same of the square-wave characterization shown in Figure~\ref{fig:tresponce}\textbf{(a)}. The excitation consists in a $V_{AC}$ sine-wave with amplitude in the range between $1.0$ and $3.5$ V summed with a $V_{DC}=11$ V DC voltage bias shown in Figure~\ref{fig:VSin}\textbf{(b)}. The bias current for this experiment was chosen to be $I_B=72\ \mu$A in order to have a sharp super-to-normal transition and a linear dependence between $R$ and $V_G$. The continuous variation of the gate voltage provides a continuous variation of the junction resistance accordingly with the $R(V_G)$ curves \cite{Puglia2020a}. Due to the former behavior, the system lies in the zero-voltage state for $I_S\left[V_G(t)\right]>I_B$, and a voltage drop across the junction appears when the condition $I_S\left[V_G(t)\right]\simeq I_B$ is satisfied. When $I_S\left[V_G(t)\right]<I_B$, the voltage drop across the nanojunction increases because of the gate-driven evolution of the constriction resistance and, eventually, reaches the saturation level at the asymptotic value of the resistance of the normal state. Figure~\ref{fig:VSin}\textbf{(a)} displays the voltage drop across the junction as a function of both the bias current and the gate voltage. The transition edge is highlighted by a dashed red line. The working point set by $I_B$ and $V_{DC}$ is shown by white dashed lines. The time resolved $V(t)$s for $V_{AC}$ equal to $1.0$ and $3.5$ V are reported in Figure~\ref{fig:VSin}\textbf{(c)}, \textbf{(d)}. Notably, the rectification threshold and the portion of the input signal rectified can be selected setting both $I_B$ and $V_{AC}$.

The former characteristics, typical of a half-wave rectifier, here are realized for the first time exploiting an all-metallic gated superconducting Dayem bridge. We speculate that the described rectifying behavior can be suitably exploited to rectify an incoming radiation coupled to the gate through an antenna, realizing a gate-controlled version of a transition edge sensor \cite{Ullom2015,Paolucci2021,Goltsman2001}. This device could operate in an extremely wide frequency range, spanning from below 1 GHz to about 1 THz. This interval is particularly relevant for both technological applications and for fundamental research (e.g. in astrophysics for cosmic microwave background detection). 

\subsubsection{Amplification properties}

The vanadium Dayem bridge, thanks to the peculiar $R(V_G)$ characteristics, is suitable for the realization of an amplifier. The gain parameter of a gated DB device is defined as the ratio between peak-to-peak amplitudes of gate voltage input and output voltage drop across the junction $g=\frac{V_{out}}{V_{in}}$. For our system $V_{out}=R(V_G)I_B$ is the voltage drop across the junction and is directly proportional to the resistance and the current bias. The $V_{in}$ is obtained by the ratio of the width of SCPD \cite{Puglia2020} and the transcoductance $\tau=\frac{dI_S}{dV_G}$. For the devices taken into account in this section $g\sim7$ with a typical power consumption of $\sim40$ nW. It is worth to highlight, on the one hand, that $g$ is of the same order of magnitude of the conventional semiconductor cold amplifier \cite{Ivanov2011,Oukhanski2003}. Such a result, on the other hand, was obtained with a power consumption of about three order of magnitude smaller than the typical semiconducting counterpart. Furthermore, a chain of $N$ rectifiers can be obtained by deliver the gate electrode of the n-th device with the output voltage of the ($n-1$)-th DB, obtaining a total gain equal to $g^N$.

The possibility to tune the critical supercurrent via conventional gating paves the way to a wide range of applications. Indeed, gate-controlled devices could be exploited to realize tunable magnetometers \cite{Clarke2004,Giazotto2010} and heat control systems \cite{Giazotto2012,Fornieri2017}. Furthermore, by exploiting the gating effect a voltage-controlled version of the nanocryotron \cite{McCaughan2018,McCaughan2014} can be implemented. The latter is a three terminal superconducting device in which a localized switching-current suppression (triggered  by injecting a control-current which generates a localized high temperature region, an hotspot, by Joule heating) destroys the superconducting properties of the nearby system. The gated version of the nancryotron, the so-called (EF-Tron) \cite{Paolucci2019a}, is implemented through the parallel of a resistor and a gated superconductor. Differently from current-driven devices \cite{McCaughan2014,Morpurgo1998}, the EF-tron is expected not to be limited by the characteristic time scale of thermal effects, that does not allows to use signals with frequencies larger than about hundreds of MHz at cryogenic temperatures \cite{Giazotto2006}. In this view, it is worth to discuss the role of an eventual direct power injection into the gated device (driven, e.g., by a gate-superconductor leakage current) that could produce an increase of the electronic temperature, detrimental for device performance.

\section{Non-thermal origin of supercurrent suppression in gated all-metallic superconducting devices}\label{sec:thermal}

\subsection{SCPDs in a titanium gate-controlled transistor}

In a current biased JJ the transition between the superconducting and the normal state for fixed values of external parameters, e.g. temperature, electric and magnetic field, is triggered by a phase slip events, i.e., local random $2\pi$ jumps of the macroscopic quantum phase $\phi$ \cite{Bezryadin2012}. The accidental nature of such events leads to a non-univocal definition of the switching current, whose value is distributed according to the switching current probability distribution (SCPD). The investigation of the SCPD of a JJ is, therefore exploited to probe dynamics of the phase slips. Here, we discuss an experiment, where a well-established technique is adopted to probe the impact of gate voltage on the frequency of phase slips events in gate-controlled metallic titanium (Ti) Josephson weak-links, with the conventional theory \cite{Kurkijarvi1972,Bezryadin2010,Fulton1972}. 

\begin{figure}[!ht]
\centering
\includegraphics[width=13.4 cm]{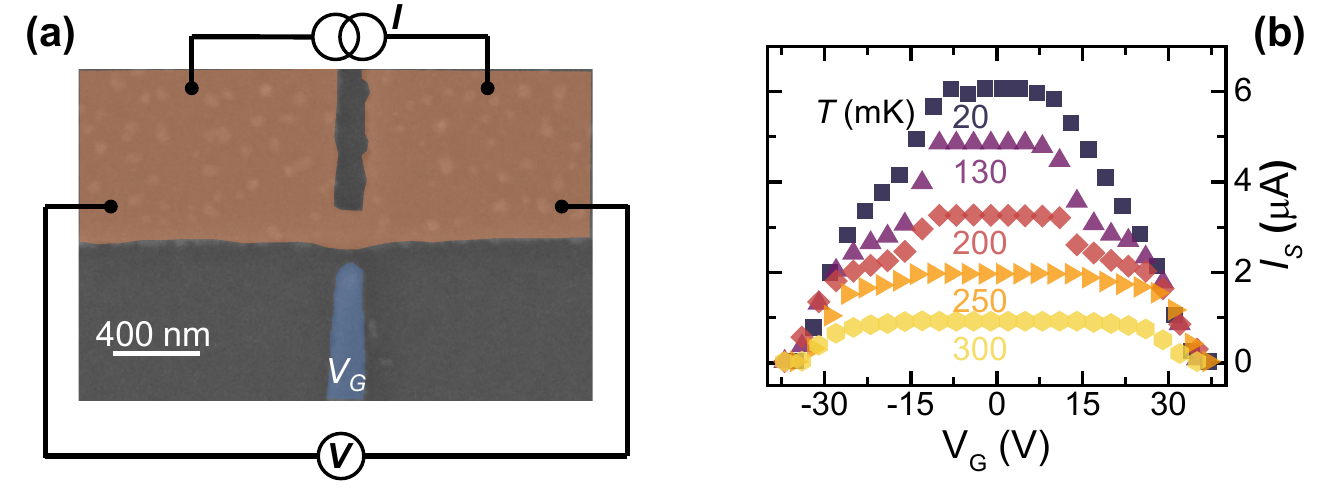}
\caption{\textbf{(a)} Pseudo-color scanning electron micrograph and bias scheme of a representative Ti gate controlled transistor. The superconducting wire and the Dayem bridge constriction are colored in orange and the gate electrode in blue. \textbf{(b)} Evolution of the switching current $I_S$ as a function the gate voltage. The switching current were acquired 50 times in order to obtain the average value shown in the picture.}
\label{fig:devscpd}
\end{figure}

The device chosen to study the evolution if the SCPD under gating action consists of a titanium Dayem-bridge. Such JJs consist of $30$-nm-thick, $10$-$\mu$m-long, $2.5$-$\mu$m-wide wire interrupted by a constriction. This $30$-nm-thick, $150$-nm-long, $120$-nm-wide narrow structure was aligned with a planar, $80$-nm-far, $140$-nm-wide metallic gate. The sample was fabricated by means of single step electron beam lithography on a single-crystal sapphire (Al$_2$O$_3$) with a typical resistivity larger than $10^{10}\ \Omega$/cm. The metal was deposited by means of an electron beam evaporation at a rate of $1.2$ nm/s. Figure~\ref{fig:devscpd}\textbf{(a)} shows a pseudo-color scanning electron micrograph.

The device shows a normal state resistance $R_{N} \simeq550\ \Omega$, a switching current at 20 mK of $I_S=6.0\ \mu$A and a critical temperature $T_{C} \simeq310$ mK \cite{Puglia2020}.

The dependence of $I_S$ \cite{Paolucci2018,Paolucci2019,Paolucci2019a,Paolucci2019b,DeSimoni2018,DeSimoni2019,DeSimoni2020,Puglia2020a,Puglia2020} on the gate voltage is showed acquiring the $I_S$ vs $V_G$ characteristics as a function of bath temperature. Figure~\ref{fig:devscpd}\textbf{(b)} shows that supercurrent vanishes for $\left|V_G\right|\simeq34$ V and such value appears to be independent from the bath temperature. By increasing the values of the temperature $I_S^0 = I_S(V_G = 0)$  lowers and a greater ineffectiveness range of the gate voltage on $I_S$ is observed. This latter behavior is the same obtained on Ti and Al superconducting gate-controlled devices \cite{Paolucci2018,DeSimoni2018}.

To characterize the effect of the temperature on a superconducting Dayem-bridges JJ, SCPDs were measured at different values of the thermal bath temperature. The distributions were reconstructed drawing an histogram based on $10^4$ switching current acquisitions for each value of the bath temperature $T$.

\begin{figure}[!ht]
\centering
\includegraphics[width=13.4 cm]{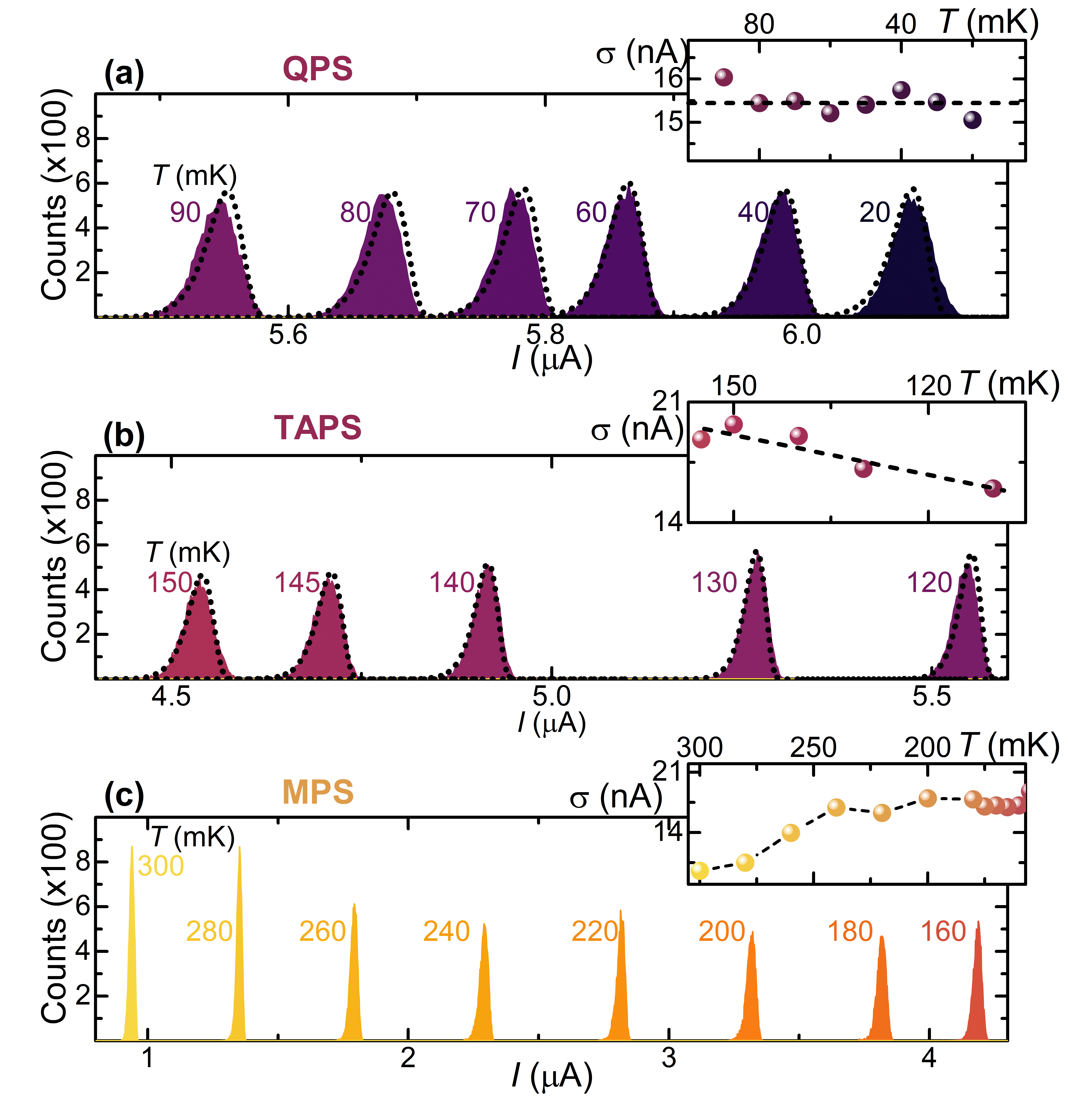}
\caption{\textbf{(a)} SCPDs vs current $I$ in the QPS regime  for temperatures between $ 20 $ and $ 90 $ mK. The inset shows the evolution of the $\sigma$ vs $T$ characteristic in the QPS regime. \textbf{(b)} SCPDs vs current $I$ in the TAPS regime  for temperatures between $ 120 $ and $ 150 $ mK. The inset shows the evolution of the $\sigma$ vs $T$ characteristic in the TAPS regime. \textbf{(c)} SCPDs vs current $I$ in the MPS regime  for temperatures between $ 160 $ and $ 300 $ mK. The inset shows the evolution of the $\sigma$ vs $T$ characteristic in the MPS regime. For each SCPD the total sampling number of $ I_S $ is $ 10^4 $. Dotted lines represent the best-fit curves obtained with KFD model.}
\label{fig:scpdthermal}
\end{figure}

Figure~\ref{fig:scpdthermal}\textbf{(a)}, \textbf{(d)}, \textbf{(c)} shows the evolution of these so-called thermal-SCPDs in a bath temperature range between $20$ and $300$ mK. The dependence of the shape of thermal-SCPDs is analyzed through the conventional Kurkijärvi-Fulton-Dunkleberger (KFD) theory \cite{Kurkijarvi1972,Fulton1971} with a fit procedure \cite{Puglia2020}. First of all the different phase slip regimes were identified thanks to the evolution of distribution standard deviation $\sigma$ as a function of $T$. The evolution of the $\sigma$ vs $T$ curve \cite{Bezryadin2012,Puglia2020} is flat at low temperature in the Quantum Phase Slip (QPS) regime, linear in $T$ in Thermal Activated Phase Slip (TAPS) regime, and decreasing for the Multiple Phase Slip (MPS) regime \cite{Bezryadin2000,Bezryadin2010,Bezryadin2012,Giordano1988,Giordano1989,Kurkijarvi1972,Fulton1974,Fulton1971}. The temperature $T_Q$, separating the QPS and the TAPS regime occurs at about $T\simeq110$ mK while the crossover between TAPS and MPS regimes appears for $T_M\simeq160$ mK.

\begin{figure}[!ht]
\centering
\includegraphics[width=13.4 cm]{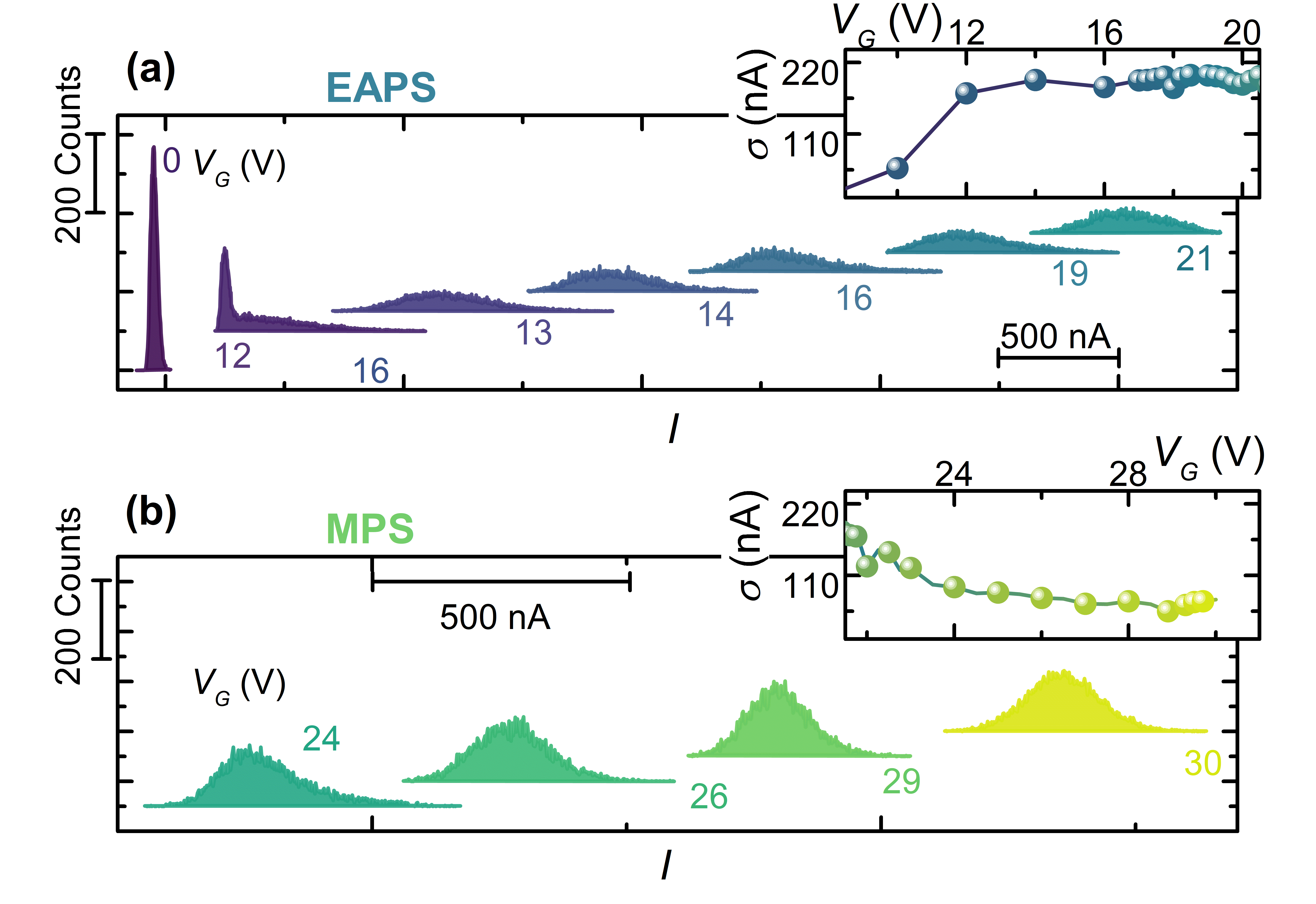}
\caption{\textbf{(a)} SCPDs vs current $I$  in the EAPS regime with gate voltages ranging from $ 0 $ V to $ 21 $ V. The inset shows the evolution of the $\sigma$ vs $V_G$ characteristic in the EAPS regime. \textbf{(b)} SCPDs vs current $I$  in the MPS regime with gate voltages ranging from $ 24 $ V to $ 30 $ V. The inset shows the evolution of the $ \sigma $ vs $ V_G $ characteristic in the MPS regime. For each SCPD the total sampling number of $I_S$ is $10^4$. The curves are horizontally and vertically offset for clarity.}
\label{fig:scpdelectric}
\end{figure}

Although these devices show the conventional evolution of the phase slips dynamic as a function of the temperature, the gate voltage drives the junction in a different regime. Figure~\ref{fig:scpdelectric}\textbf{(a)} shows vertically shifted SCPDs collected for several values of the gate voltage at $ T=20 $ mK. In particular, the SCPDs are overlapping for $ V_G < 8 $ V, and a low current tail appears for $8<V_G<14$ V. In addition, the distributions greatly widens for $14<V_G<24$ V and for high gate voltage values, i.e. $V_G>24$ V, the SCPDs turns out to narrow. In this electrostatically driven scenario, the $\sigma$ vs $V_G$ curve takes the place of the conventional $\sigma$ vs $T$ characteristic. As shown in Figure~\ref{fig:scpdelectric}\textbf{(b)}, $\sigma(V_G)$ curve exhibits  a region where $\sigma$ is constant, thereby showing a marginal effect of the gate voltage to the phase slips for small $V_G$ values. This behavior is  similar to the QPS regime. Therefore, we identify a  crossover gate voltage $V_Q\simeq8$ V between the former and the Electric Activated Phase Slip (EAPS) regime, where the $\sigma$ grows up to $\sim200\ nA$ as the gate voltage increases. Notably, $\sigma$ starts to increase at the same voltage at which the switching current begins to be affected by the electric field. Such evidence seems to suggest that, regardless of the  origin of $I_S$ suppression, the latter is associated with a corresponding raise of the number of phase slip events. Finally, for greater values of the gate voltage (i.e.,  $V_G>V_E\simeq14$  V), the standard deviation$\sigma$ decreases and saturates to $\sim75\ nA$.

\begin{figure}[!ht]
\centering
\includegraphics[width=13.4 cm]{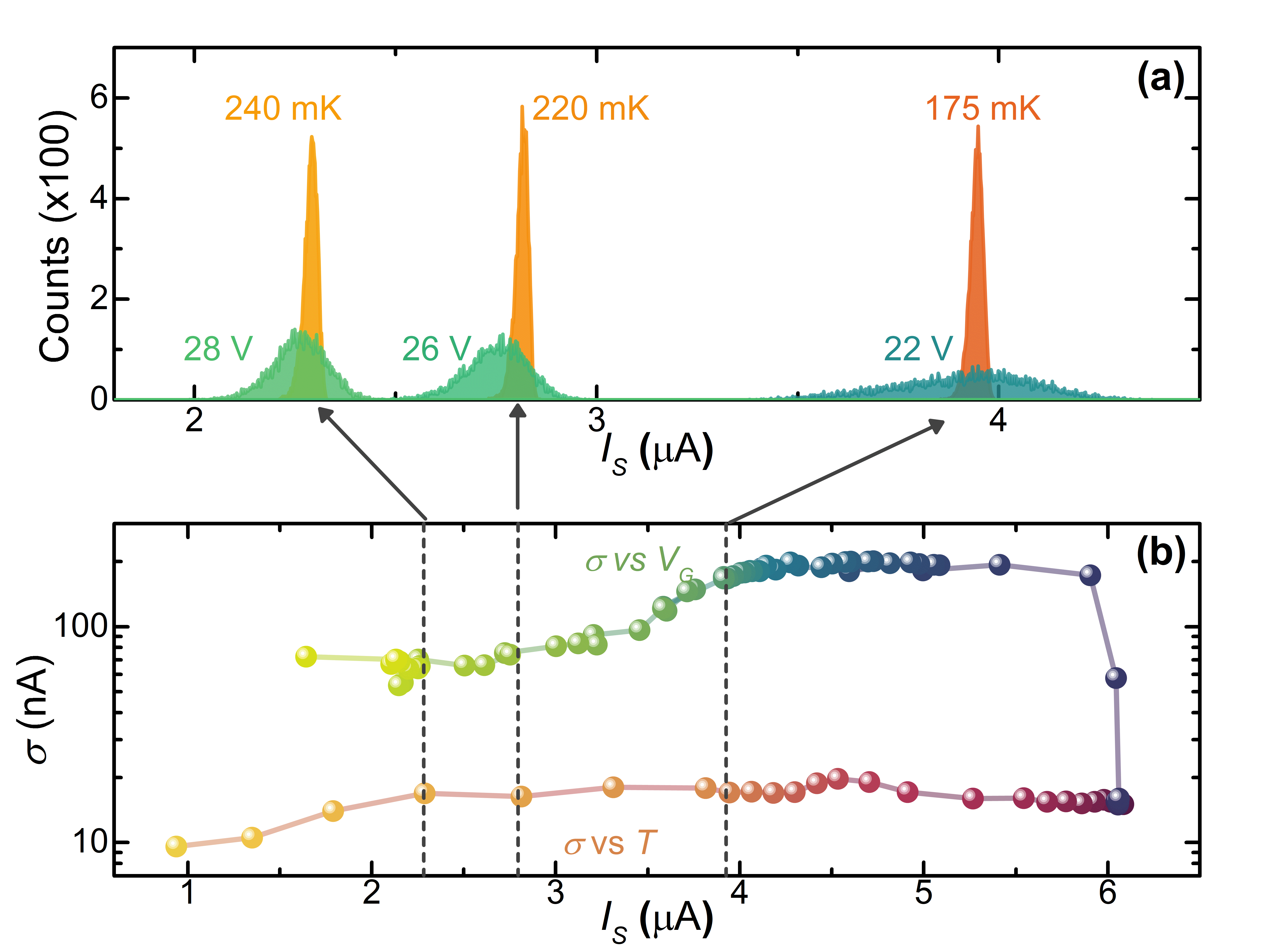}
\caption{\textbf{(a)} $I_S$-matched SCPDs. Red/orange SCPDs were acquired with a negligible electric field $ V_G=0 $ V at chosen temperatures and blue/green SCPDs were obtained at a bath temperature of $ T = 20 $ mK for different gate voltages. The values of $I_S$ are respectively from left to right $2.2,\ 2.8,\ 4.0\ \mu$A. \textbf{(b)} $\sigma$ vs $I_S$ characteristics obtained for thermal (lower curve) and electric (upper curve) distributions.}
\label{fig:scpdcomparison}
\end{figure}

The starkly different behavior between thermal- and electric- SCPDs is clearly displayed in Figure~\ref{fig:scpdcomparison}\textbf{(a)} where three $I_S$-matched couples of thermal and electric distributions are plotted in the same graph for comparison. The $I_S$-paired SCPDs highlight remarkably different widths and shapes. Such behavior seems to stem from a gate-driven strong non-equilibrium state induced in the junction. Concerning the standard deviation of the distributions, on the one hand, the comparison between the $\sigma$ vs $I_S$ curves extracted from the two thermal- and electric- SCPDs series, shown in Figure~\ref{fig:scpdcomparison}\textbf{(b)}, displays a qualitatively-similar behavior. On the other hand, the electric driven SCPDs present a $\sigma$ on average around one order of magnitude larger.

Indeed, if we assume that the voltage-driven broadening of the SCPD is due to an increase of the electronic temperature, e.g. a trivial Joule heating due to a gate-DB current, we run into the absurdity of obtaining an electronic temperature higher than the critical temperature of the superconductor\cite{Giazotto2006,Bezryadin2012}. This observation reflects into the impossibility to fit the gate-activated distributions by means of a conventional KFD transform, since the obtained parameters would be nonphysical. Therefore, these data demonstrate on the one hand a strong link between phase slip events and electric field. On the other hand, they suggest a non-thermal origin of the switching current suppression: the action of the gate voltage drives the DB in a state whose description is incompatible with that of a superconductor heated through a voltage-driven power injection at a thermal steady state with electronic temperature higher than that of the thermal bath.

\subsection{Suspended titanium gate-controlled transistor}

We have shown that it is not possible to describe the modification of the SCPDs as a consequence of a trivial overheating. In other words, the effect of the gate is unlikely that of driving the superconductor into a higher electronic-temperature steady state by a mere power injection (driven by a current injection). To further investigate the role of a possible injection of current between the gate and superconducting channel, fully-suspended gated superconducting nanobridges were tested. In  conventional gated devices two gate-channel charge transport mechanisms might be present: the diffusive current injection through the substrate and the ballistic cold-electron field-emission (CFE) through the vacuum. The suspended geometry permits to exclude the first leaving the CFE as the only possible charge transport mechanism.

\begin{figure}[!ht]
\centering
\includegraphics[width=13.4 cm]{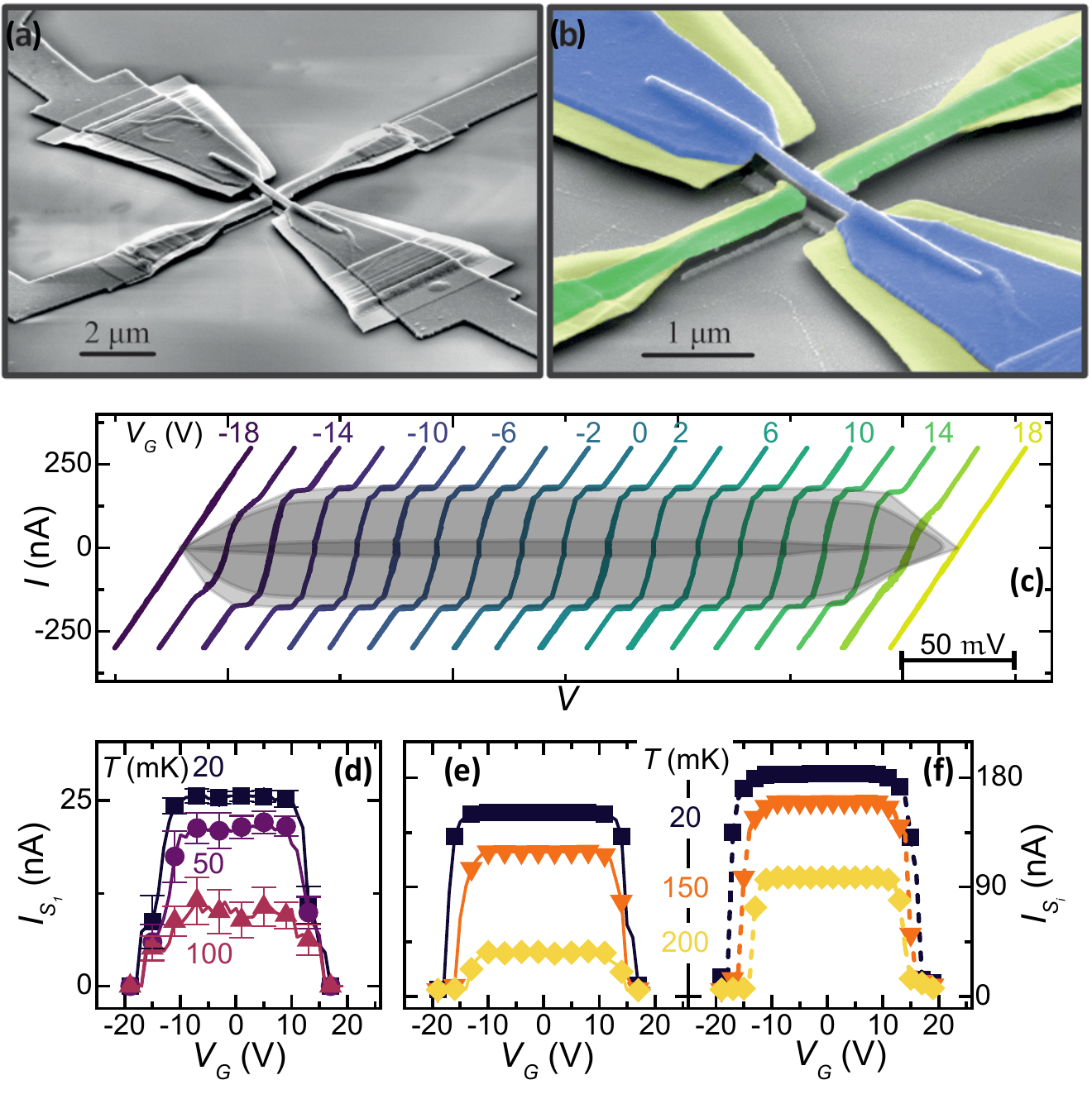}
\caption{\textbf{(a)} SEM image of a representative device. \textbf{(b)} The wire (blue) is connected in a typical four-wires configuration. The switching current $I_S$ was decreased via the application of the same gate voltage $V_G = V_{S_1} = V_{S_2}$ to the symmetric gate electrodes (green). \textbf{(c)} Back and forth current $I$ vs  voltage $V$ curves for selected gate voltage $V_G$ values acquired with a bath temperature of $ T = 20 $ mK. The curves are horizontally shifted for clarity. Grey regions highlight the decrease of the switching currents $I_{S_1}$, $I_{S_2}$ and $I_{S_3}$. \textbf{(d)} $I_{S_1}$ vs gate voltage $V_G$ acquired at different bath temperatures $ T $. \textbf{(e)} and \textbf{(f)} show the gate voltage $V_G$ dependence of $I_{S_2}$ and $I_{S_3}$, respectively. Error bars represents the standard deviation of $ I_{S_i} $ with $i=1,\ 2,\ 3$ calculated over $25$ repetitions.}
\label{fig:susdev}
\end{figure}

This experiment was performed on titanium gated suspended wires. The devices consist of $70$-nm-thick  and 1.7-$\mu$m-long nanobridge deposited on top of an undoped $130$-nm-suspended crystalline InAs \cite{Li2014,Iorio2019} nanowire set on two Ti/Au-coated ($5/15$ nm) pillars of cross-linked insulating PMMA. The wires were gated through two $\sim$350-nm-wide, $40$-nm-far symmetric Ti side electrodes. Titanium was evaporated at a rate of about 1.2 nm/s. Figure~\ref{fig:susdev}\textbf{(a)},\textbf{(b)} shows a 3D-sketch and a pseudo-color scanning electron micrograph of a typical device.

Such a suspended nanojunction show four different superconducting transition \cite{Rocci2020}, that can be interpreted as the switch of the superconducting banks for $I_B\simeq1.8\ \mu$A and of the series of three junction with switching current respectively $I_{S_1}\simeq25\ nA$, $I_{S_2}\simeq150\ nA$ and $I_{S_3}\simeq180\ nA$. The geometry of the junction and the multi-step fabrication process induced to the existence of such series of three junction in the nanobridge due to inhomogeneities  of the titanium layer thickness, covering a InAs nanowire. Moreover, the switching current difference resides in a the variation of the cross-section of Ti film coating the wire and on the inhomogeneous anti-proximization effect of the superconducting film due to the bottom gold layer.

$I$ vs $V$ shifted characteristics at selected gate voltage from $-20$ to $20$ V of the bridge  are shown in Figure~\ref{fig:susdev}\textbf{(c)} at a temperature of 20 mK. Figure~\ref{fig:susdev}\textbf{(d)},\textbf{(e)},\textbf{(f)} shows the evolution of the switching currents $I_{S_i}$ of the three junctions as a function of the voltage gate, extracted from the $I$ vs $V$ curves measured at several bath temperatures.

Notably, as the temperature increases, the ineffectiveness plateau shrinks. The former observations is in starkly contrast with previous works \cite{DeSimoni2018,Paolucci2018,Paolucci2019}, in which the plateaux broadened and the quenching gate voltages were constant as the temperatue increases. We attribute such discrepancy to a smaller substrate-to-bridge thermal coupling compared with devices located on a  substrate. Certainly, independently  of the microscopic origin of the gate  effect, the quenching of the switching current could be connected to a sizable increase of the quasi-particles number in the superconducting nanowire \cite{Puglia2020,Alegria2021}. Such enlargement is expected to be more efficient in a suspended geometry  in which the relaxation of quasiparticle excitations via  electron-phonon interaction is reduced compared to conventional devices. These results demonstrate that the presence of an interface between the substrate and the superconducting junction is not necessary for the gate effect to occur. This is a unequivocal  evidence against the hypothesis of a Joule heating origin of the supercurrent suppression due to a diffusive current injected into the substrate.

\subsection{Leakage current finite element method simulations}

The suspended geometry experiment allows to eliminate any Joule overheating conveyed to the nanowire through phonon coupling due to a  leakage current injected via the  substrate. In this framework, a current flowing from the bridge to the gate (and vice versa) could be possible only via the field emission of cold electrons (CFE) in the vacuum. Such emission is expected to arise because of the application of a strong electrostatic field between the nanowire and the gate electrodes \cite{Simmons1963,Fowler1928}. 

\begin{figure}[!ht]
\centering
\includegraphics[width=13.4 cm]{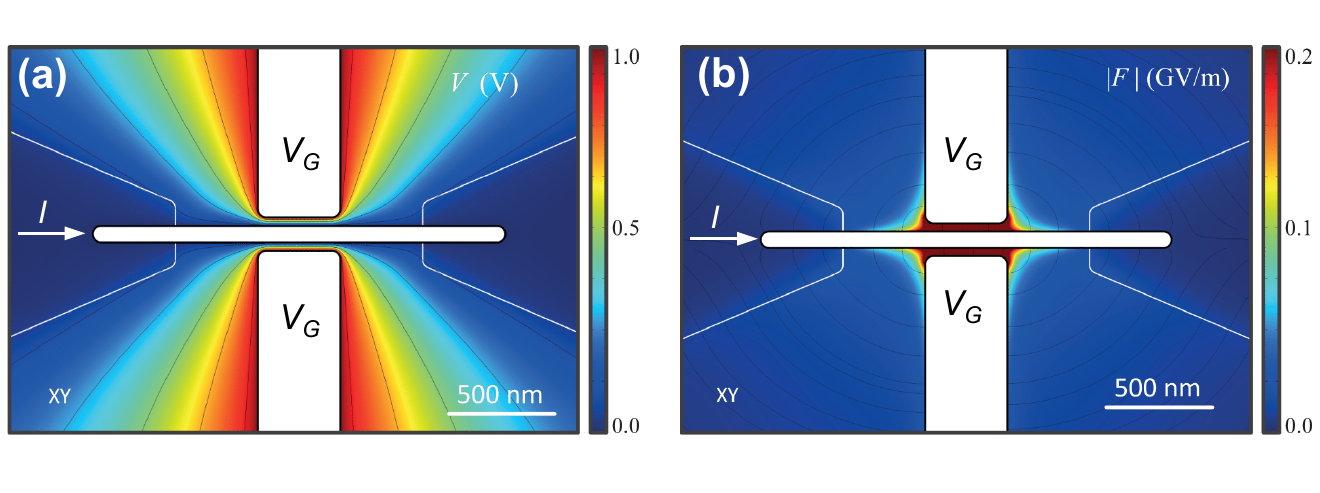}
\caption{\textbf{(a)} Electric potential $V$ in a XY cut-plane which intersects the suspended nanowire at half height. Applied gate voltage $V_G = 15$ V. \textbf{(b)} Color-plot showing the derived electric field amplitude $|F|$ distribution from the solution of the electric potential. Values obtained for $V_G= -15 $ V. The simulation shows that the electric field is extremely localized.}
\label{fig:simule}
\end{figure}

To investigate the role of a possible field-emitted  current in the $I_S$ suppression, the cold emitted current ($I_{FE}$) can be quantified with a  3-dimensional finite-element  method simulations implemented on the same geometry of the suspended titanium gate-controlled transistor, and then compared  with the measured leakage current $I_L$ \cite{Rocci2020}. $I_{FE}$ is calculated via the integration over the surface of the cathode, i.e., the gate (wire) for negative (positive) $V_G$, of the Fowler Nordheim tunnel  current density that at the cathode is written as \cite{Fowler1928,Simmons1963}

 $$  \vec{J}_{FE}(\vec{F}, h_0) = \frac{2.2 e^3 }{8 \pi h h_0} \vec{F}^2 \exp \left[ - \frac{8 \pi }{2.96 h e |\vec{F}| } (2 m_e)^{1/2} h_0^{3/2} \right],$$

where $\vec{F}(x,y,z)$ is the modulus of the electrostatic field at the surface of the cathode, $m_e$ is the electron mass, $e$ is the electron charge, $h$ is the Plank’s  constant, and $h_0 = 4.3$ eV is the work function of titanium \cite{Wilson1966}. The electrostatic field vector $\vec{F}(x, y, z;  V_G)$ was computed in the 3D vacuum space zone, around the  bridge and the side gates, via the Maxwell equation $\vec{F} = -\nabla V(x, y,z; V_G)$. The electric potential $V (x, y,z; V_G)$ was calculated thanks to the numerical integration of the Poisson equation $\nabla^2 V(x, y, z;  V_G) =  0$ (see Figure~\ref{fig:simule}\textbf{(a)}). As boundary condition, the bridge and the gate  electrode surfaces were simulated with perfect  equipotential conductor at $V = 0$ and $V = V_G$, respectively.

The propagation of the electric field module $\vec{|E|}$ in the space obtained for the entire 3D domain of the simulation is color-plotted in Figure~\ref{fig:simule}\textbf{(b)} from a top view ( X-Y ) for $V_G  =  -15$ V. The electrostatic field is confined in the region between the titanium bridge and the lateral gate surfaces and it quickly fades elsewhere, so not influencing the superconducting banks.

The electric field reaches the maximum intensity of $0.2\ GV/m$ in correspondence of the center of the gate and it is localized near the side gate surfaces. Moreover, $|\vec{F}(x,y,z)|$ drops more than one order of magnitude between $500$ nm from the lateral edge of the gate electrode. Such a field geometry let us to conclude that the banks are unlikely to be affected by the gate voltage.

\begin{figure}[!ht]
\centering
\includegraphics[width=13.4 cm]{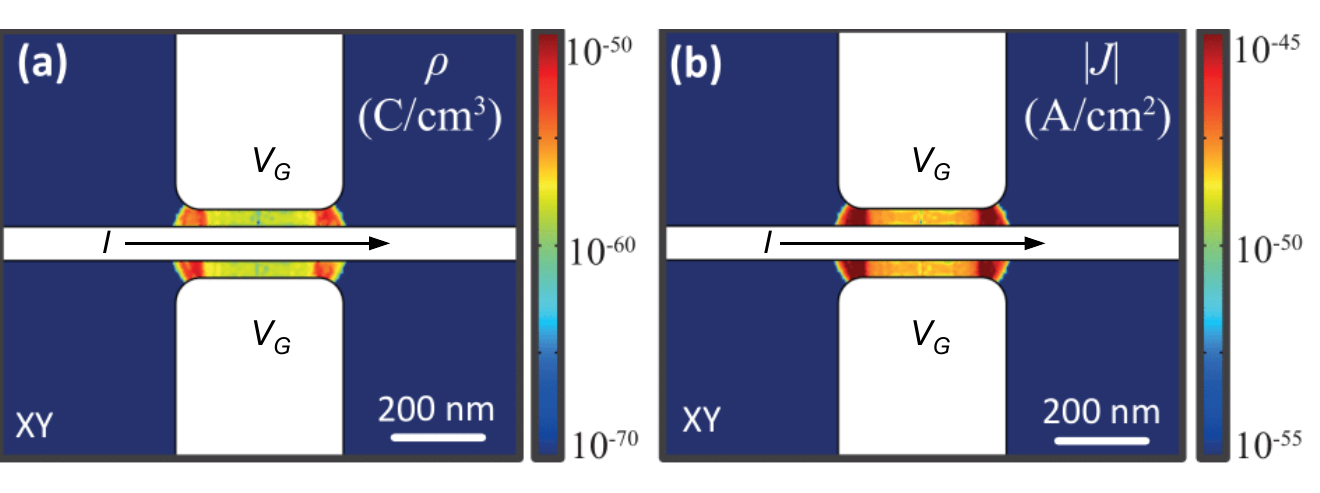}
\caption{\textbf{a} Color-plot of the induced charge density $ \rho $. \textbf{(b)} Modulus of the propagating current density $|J|$obtained solving the ballistic trajectories of the emitted electrons. Values obtained for $\phi_0 = 4.3 $ eV and $ V_G = -15$ V. The distribution of the current shows that the injection mechanism influences only a $ 400 $ nm section of the nano-bridge.}
\label{fig:Jfe}
\end{figure}

By solving the ballistic trajectories of the electron emitted by the electrode, it is possible to compute the current density $|\vec{J}_{FE}|$ in the region between the gate and the wire. Figure~\ref{fig:Jfe}\textbf{(a)}, \textbf{(b)} shows the induces charge density $\rho$ and the current density $|\vec{J}_{FE}|$ evaluated on a X-Y plane. It is worth to notice the extreme localization of the electrons in a region of about $500$ nm centered on the electrodes that affect only a small portion of the nanobridge. Such evidence prove that for CFE the the entire number of the electrons emitted/absorbed from the wire is absorbed/emitted from the gate.

\begin{figure}[!ht]
\centering
\includegraphics[width=13.4 cm]{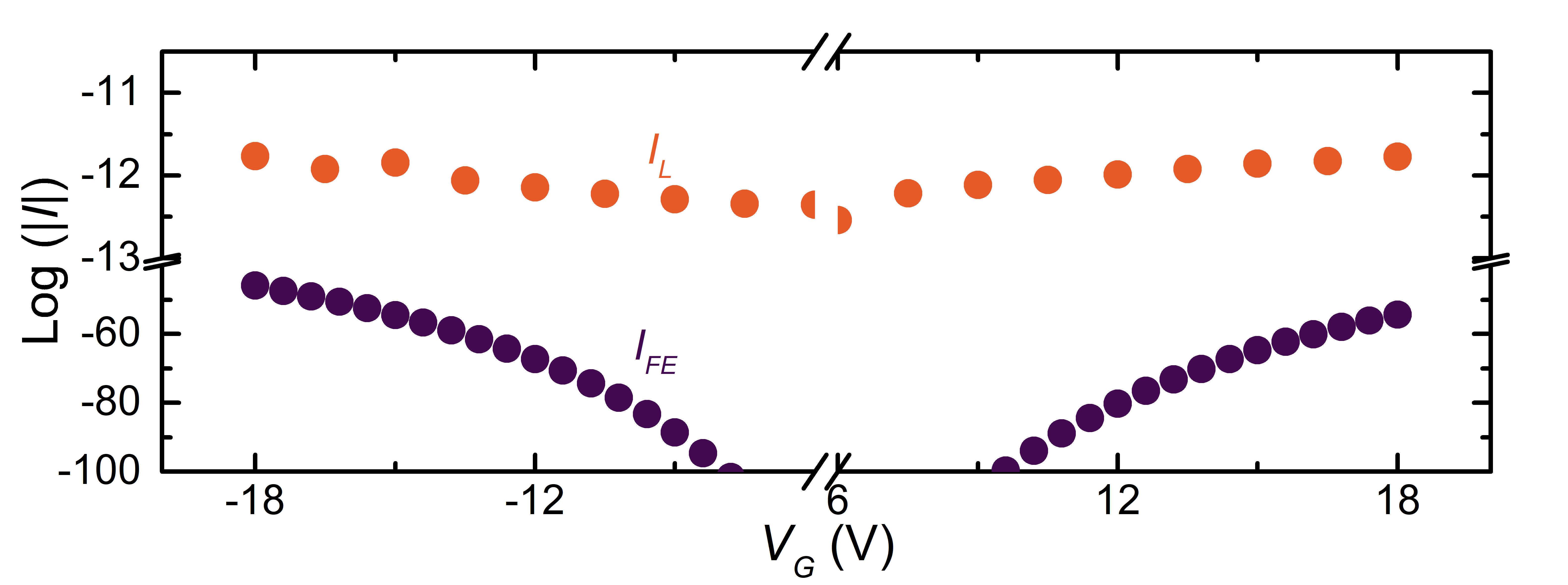}
\caption{Comparison between the measured leakage current $ I_L $ (orange points) and the field emission current $\left(I_{FE}\right)$ (violet points) as a function of the gate voltage. The latter quantity wad obtained integrating numerically the Fowler-Nordheim current density $ \vec{J}_{FE} $ with the titanium work function $ \phi_0 = 4.3 $ eV.}
\label{fig:simulIleak}
\end{figure}

Finally, the integration of the current density $\vec{J}_{FE}$ over the surface of the electrodes returns the current $I_{FE}$, a quantity that can be directly compared with the gate-wire current measured in the experiment (shown in Figure~\ref{fig:simulIleak}\textbf{(a)}) \cite{Rocci2020}. Figure~\ref{fig:simulIleak}\textbf{(b)} shows the results of the simulation for $I_{FE}$. 

Notably, $I_{FE}$ is more than $20$ orders of magnitude smaller than the greatest value of the gate-bridge $I_L$ measured in the experiment. Furthermore, a current of about $10^{-40}$ A corresponds to the emission, on average, of a single electron in $10^{28 }$ years. Such emission rate was computed consistently with an electrostatic field at the surface of the that is not intense enough to start a true CFE current. As a matter of fact, cold-emitted electrons typically needs an electrostatic field of the order of $1 - 10$  GV/m \cite{Bhushan2012}, but in our situation the greatest electrostatic field is one order of magnitude smaller, at most. Moreover, the simulation shows an intrinsic asymmetry, of several orders of magnitude, of $I_{FE}$ when $V_G\longrightarrow-V_G$ due to the geometry difference between gate and wire. This seems to suggest that, if the field emitted current was the leading mechanism in determining the $I_S$ suppression, a strongly asymmetric behavior should be observed for positive and negative gate voltages. Such a feature was never reported in experiments on gated metallic superconductors \cite{DeSimoni2018, Paolucci2018,Paolucci2019}.

\subsection{Heating through single cold-electron field emission or absorption}

If we admit that a certain number of electrons are emitted or absorbed by the gate and absorbed/emitted by the wire the expected experimental  phenomenology should be totally different compared to those observed: an electron with an energy of the order of $10$ eV, and ballistically hitting the junction going through the vacuum, is expected to release its entire energy increasing abruptly the system electronic  temperature. A straightforward calculation for the heat capacitance due to electronic contribution $C_e$ of a weak-link in the normal-state is:

$$C_e=\Omega\gamma T_e$$

where $\Omega$ is the volume occupied by the junction, $\gamma$ is the Sommerfeld  constant for titanium, $T_e$ is the electronic temperature of the weak-link. 

The energy $E (V) $ released in the junction is directly proportional to the acceleration gate voltage $V$ between the weak-link and the gate electrode:

$$E\left( V \right)=q V, \ \ \ \ \ \ P\left( t \right)=E\ \delta \left( t \right)$$

where $q$ is the electron charge, $\delta$ is Dirac’s delta distribution and $P( t )$ is the power as a function of $t$. According to conventional theory of the heat transport, the electronic temperature in the system is conveniently described by the differential equation \cite{Giazotto2006}, where $T_B$ is the lattice temperature:

$$C_e\frac{\partial T_e}{\partial t}=P\left(t\right)\longrightarrow T_e=\sqrt{\frac{2E}{\Omega\gamma}+T_B^2}$$

We wish to stress that by this approach we obtain  final electronic temperature $T_e$ which is an underestimate of its real value in the weak-link since we considered $C_e$ to be the normal-state heat capacitance, that is generally exponentially greater than in the superconducting state \cite{Giazotto2006}. The, already mentioned, calculation demonstrates that one electron accelerated with a $30$ V gate voltage and injected into a typical DB constriction at $10$ mK of bath temperature would increase starkly its electronic temperature $T_e$ that would reach a value that is more than $20$ times greater than its critical temperature ($T_C\simeq500$  mK). Such a conclusion permits us to make the statement that the heat generating from the absorption of a field-emitted electron in the junction in incompatible with an equilibrium state with a definite gate-tunable $T_e$ and $I_S$. 

\subsection{Continuous power injection}

\begin{figure}[!ht]
\centering
\includegraphics[width=13.4 cm]{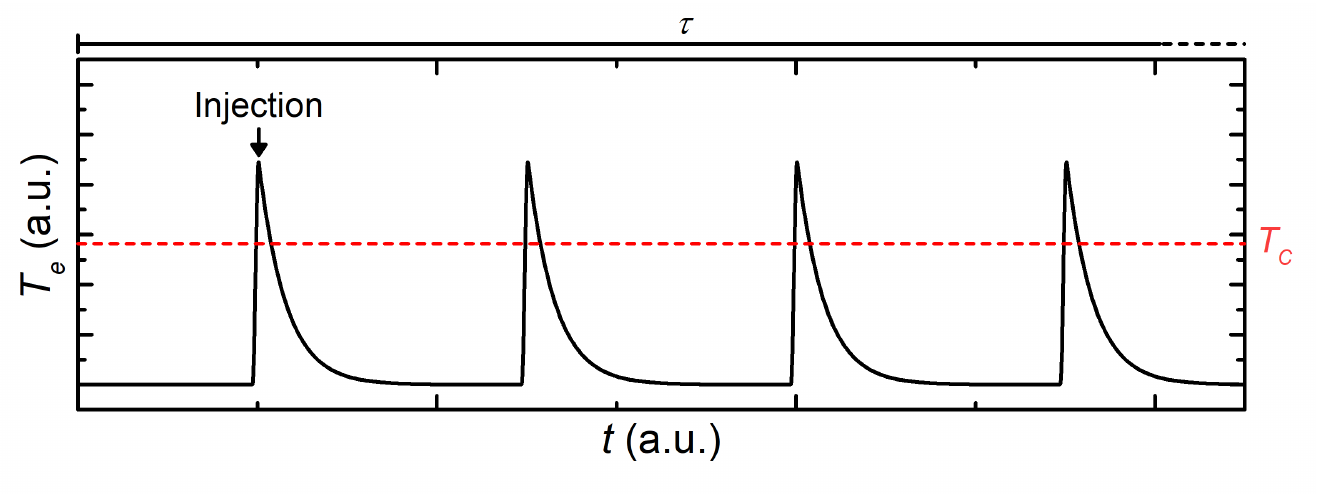}
\caption{Electronic temperature $T$ vs time $t$ of a mesoscopic superconducting weak-link that periodically absorb electrons with energy of the order 10 eV. The red horizontal line represents the critical temperature of the superconductor. Each electron starkly increase the electronic temperature of the system driving it in the normal state. $\tau$ is the measurement time.}
\label{fig:dynamic}
\end{figure}

In principle, we can assume that, the repeated absorption of high-energy electrons, leads the system to bouncing continuously between the normal and superconducting states as shown in Figure~\ref{fig:dynamic}. 
If this were the case, the bridge is suddenly driven into the dissipative state  for every electron absorbed. Then, the system relaxes back to the dissipationless state. The recurrence of such events is due to the emission rate of the electrons, and the super-to-normal relaxation time is typically given  by the electron-phonon coupling related time, that is of the order of $\tau\simeq1$ ns \cite{Giazotto2006}. Such time scale is much smaller than the typical  integration time of our voltage measure setup ($ \simeq 20$  ms). In this scheme, in the time of a voltage $V$ vs current $I$ measurement, every time an electron is absorbed by the junction, if the bias current $I$ is below the $I_R$ the transition that lead to a finite resistance value is likely too fast to be detected by our setup. On the contrary, if $ I > I_R $, the absorption of an electron by the constriction drives the system to the normal state, and it should persist in such state until the flowing current $I$ is lowered to $0$. Such consideration suggest that when electron are field-emitted, $I_R$ and $I_S$ should always correspond. Since this does not occur, we conclude that any hot-electron  injection-related mechanism related to field emission as the main cause of our  experimental results, should be excluded.

\subsection{Unconventional sum rule}

\begin{figure}[!ht]
\centering
\includegraphics[width=13.4 cm]{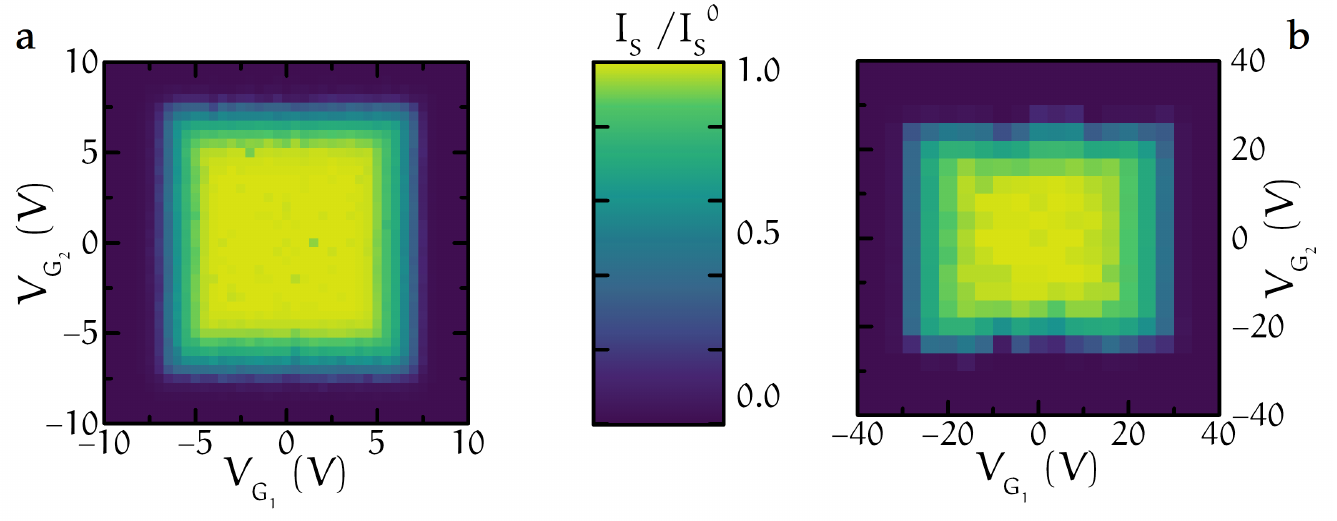}
\caption{\textbf{(a)} and \textbf{(b)}Effect of the electric field generated in a double gate flanked Dayem bridge symmetric side gate electrodes. Contour-plots show the normalized supercurrent as a function of the two gate voltages $V_{G_1}$ (x axis) and $ V_{G_2 }$ (y axis) for two representative devices (A and B).}
\label{fig:sumrule}
\end{figure}

Another strong evidence against a trivial heating or a direct power-injection origin of the supercurrent suppression comes from the evolution of the switching current of a superconducting wire under the action of a pair of lateral side gates. By means of a titanium Dayem bridge consisting of a double gate flanked nanoconstrictions interrupting a Ti strip \cite{Paolucci2019a}, it was possible to asses the mutual influence and the spatial extension of two opposite gate electrodes effect in the suppression  of the switching current. Figures~\ref{fig:sumrule}\textbf{(a)},\textbf{(b)} show two color-plot of the normalized switching current $I_S/I_S^0$ vs $V_{G_1}$ (X  axis) and $V_{G_2}$ of two representative devices $A$ and $B$. The difference of the two systems in the complete quenching gate voltage $V_G^C$ was ascribed to the different in the gate-DB distance ($ \simeq 80$ nm for the sample $A$ and $ \simeq 120 $ nm for the sample $B$). The observed square-like shape indicates the existence of a voltage threshold $V_{th}$: when one of the two gates is biased above $V_{th}$ the critical supercurrent is suppressed by a fraction which is not dependent of the voltage applied to the other gate. In other word, the effect of the two gates on $I_S$ are independent and no obvious sum rule exists between the action of the two voltages. Such evidence suggests that the gate-driven suppression of the supercurrent is likely related to a surface effect, which affects non-locally superconductivity, i.e., once that the electric field established at one of the surface of the superconductor overcomes a critical value, its effect is propagated inside the superconducting body over a distance at least comparable with the device width. Accordingly, with previous works \cite{DeSimoni2018} and theoretical calculations \cite{Ummarino2017}, the surface perturbation could affect the superconductor for a  thin depth  of the order of the superconducting coherence length $\xi$. Furthermore, the aforementioned behavior is hardly comparable with the picture of a direct heat/power injection due to charge transport from/to the gate. Indeed, in the latter case a sum rule for the total power $P_{sum}$ is expressed as $V^2_{G_1}/R_1+V^2_{G_2}/R^2$, where $R_{1,2}$ are the gate-superconductor leakage resistance respectively for the gate 1 and 2, respectively.

\section{Summary and further research}\label{sec:sum}
Along this review we showed the electrostatic control of the superconducting properties in several all-metallic Josephson weak-links: niobium and vanadium Dayem bridges and titanium suspended wires (Section \ref{sec:mat}). On the one hand, we focused on the technological application of the effect, demonstrating supercurrent suppression on the material that represents the industrial standard for superconducting electronics, the niobium. Moreover, the vanadium Dayem bridge experiments showed the potential for electric signal rectification of such geometry. On the other hand, we investigated the dynamic of the phase slip in a titanium Dayem bridge JJ under the effect of an electrostatic field. The results demonstrated that it is impossible to ascribe the modifications of the shape of the SCPDs to a conventional heating effect. In particular, in the framework of the established KFD theory, it was not possible to interpret the width of the distribution with the usual parameters. Furthermore, the experiment carried out with a suspended titanium wire demonstrated that the presence of the substrate is not critical to the occurrence of the effect. Such evidence confutes any possible contribution to the superconductivity quenching due to the existence of an injection current that flows in the substrate.

In the second part of this review, we faced the hypothesis of the thermal origin of the electrostatic effect on BCS superconductors. Thanks to finite element simulations performed on a system with the same geometry of the suspended titanium transistor, we demonstrated that the cold field emission cannot be a satisfactory explanation for the suppression of the supercurrent. In fact, even assuming that single electrons are emitted from the gate and absorbed by the junction, the local increase of the electronic temperature of the system is incompatible with the superconducting state. Additionally, the non-trivial summation rule of two side gate in the quenching of the supercurrent is a further evidence that a trivial thermal effect is no able to explain our unconventional gating effect.

To shed light on these experiments and to progress towards the understanding of the origin of the effect a set of complementary experiments are required. For example, SQUID microscopy could provide useful information on the distribution of the supercurrent in the field-affected region of the superconductor. In addition, scanning tunnelling spectroscopy and scanning gate experiments are critical to investigate the spatial variation of the superconducting gap. Moreover, radio-frequency-based experiments are crucial to acquire information on the characteristic time scale of the effect and on the role of the quasiparticle excitation in the quenching of the supercurrent. 

Besides a theoretical model able to explain the observed phenomenology has not been provided yet. From a technological point of view, the unconventional field-effect promises to be suitable for a wide range of applications. We already demonstrated the rectification properties of a Dayem bridge system and more complex devices such as gate- controlled radiation detector \cite{Zmuidzinas2004,Gousev1994,Losch2019}, signal routers and computational systems \cite{Likharev2012,Mukhanov2011} are at reach.
 
\end{paracol}

\begin{paracol}{2}
\linenumbers
\switchcolumn

%%%%%%%%%%%%%%%%%%%%%%%%%%%%%%%%%%%%%%%%%%

\vspace{6pt} 

%%%%%%%%%%%%%%%%%%%%%%%%%%%%%%%%%%%%%%%%%%
%% optional
%\supplementary{The following are available online at \linksupplementary{s1}, Figure S1: title, Table S1: title, Video S1: title.}


%%%%%%%%%%%%%%%%%%%%%%%%%%%%%%%%%%%%%%%%%%
%\authorcontributions{For research articles with several authors, a short paragraph specifying their individual contributions must be provided. The following statements should be used ``Conceptualization, X.X. and Y.Y.; methodology, X.X.; software, X.X.; validation, X.X., Y.Y. and Z.Z.; formal analysis, X.X.; investigation, X.X.; resources, X.X.; data curation, X.X.; writing---original draft preparation, X.X.; writing---review and editing, X.X.; visualization, X.X.; supervision, X.X.; project administration, X.X.; funding acquisition, Y.Y. All authors have read and agreed to the published version of the manuscript.'', please turn to the \href{http://img.mdpi.org/data/contributor-role-instruction.pdf}{CRediT taxonomy} for the term explanation. Authorship must be limited to those who have contributed substantially to the work~reported.}

\funding{The authors acknowledge the European Union’s Horizon 2020 research and innovation program under Grant Agreement No. 800923 (SUPERTED) for partial financial support.}

%\informedconsent{Please add ``Informed consent was obtained from all subjects involved in the study.'' OR ``Patient consent was waived due to REASON (please provide a detailed justification).'' OR ``Not applicable'' for studies not involving humans.}

%\acknowledgments{In this section you can acknowledge any support given which is not covered by the author contribution or funding sections. This may include administrative and technical support, or donations in kind (e.g., materials used for experiments).}

\conflictsofinterest{The authors declare no conflict of interest.}

%%%%%%%%%%%%%%%%%%%%%%%%%%%%%%%%%%%%%%%%%%
\end{paracol}
\reftitle{References}


%\externalbibliography{yes}
%\bibliography{references.bib}

\end{document}